# Title

Phen-Gen: Combining Phenotype and Genotype to Analyze Rare Disorders

# Affiliations

Asif Javed[1], Saloni Agrawal[1], Pauline C. Ng[1]

[1]Computational & Systems Biology, Genome Institute of Singapore, Agency for Science, Technology and Research, Singapore

Corresponding authors: Asif Javed (javeda@gis.a-star.edu.sg) and Pauline C. Ng (ngpc4@gis.a-star.edu.sg)

# Abstract

We introduce Phen-Gen, a method which combines patient's disease symptoms and sequencing data with prior domain knowledge to identify the causative gene(s) for rare disorders. Simulations reveal that the causal variant is ranked first in 88% cases when it is coding; which is 52% advantage over a genotype-only approach and outperforms existing methods by 13-58%. If disease etiology is unknown, the causal variant is assigned top-rank in 71% of simulations.

# Introduction

Rare disorder analysis has been facilitated by next generation sequencing. The diagnostic yield in large clinical studies remains moderate, varying between 16-50%[1,2]. The cause of two-thirds of new cases remains unknown[3] and nearly half of OMIM-reported disorders have an unknown molecular basis[4]. Issues that may be limiting the success of a study include focus on coding exome with exclusion of potential regulatory variants[5], and lack of systematic integration of prior knowledge of disease and gene(s) involved. More recently there has been a shift in paradigm to incorporate patient symptoms in the causative gene prediction; eXtasy[6], PHIVE[7] and PHEVOR[8] are recent examples. However each has its limitations and none of the existing symptom-driven methods extend across the entire genome.

We introduce Phen-Gen, a method which combines the patient's sequencing data and symptoms with prior knowledge of human diseases and functional interplay of different genes, all within a systematic Bayesian framework (Supplementary Fig. 1). The method offers two alternate paths to disease-gene implication. Phen-Gen's exome-centric approach predicts the damaging impact of coding mutations using nonsynonymous, splice site, and indel predictors within a unifying framework (Supplementary Figs. 2 and 3). This allows for a direct quantitative comparison between coding variants. Phen-Gen's genome-wide approach utilizes evolutionary conservation, ENCODE predicted functionality, and proximity to coding sequence to estimate disease functionality of a locus (Supplementary Figs. 4 and 5). Both approaches take into account that healthy humans harbor hundreds of supposedly damaging mutations and certain genes are more prone to these so-called deleterious variants (Supplementary Fig. 6).

To safeguard against overfitting, each component of our loosely coupled pipeline has been individually evaluated in cross-validation; before analyzing the complete method in totality. Phen-Gen assigns the causal gene first rank in 88% of simulations. It shows a 19-58% improvement in direct simulation comparison to previously published methods: eXtasy[6], VAAST[9], and PHEVOR[8]; and 13-16% improvement over results reported by PHIVE[7] in a comparable simulation framework. For novel genes, unknown to our internal database, Phen-Gen assigns first rank in 71% of simulations. The efficacy of Phen-Gen is further established in real patient data from a recently published study[2]. The causal gene is ranked first in 8/11 patients (and within top 5 in all). For ease of use and to address data privacy issues, Phen-Gen is available both as a web server and an open-source downloadable package (http://phen-gen.com).

# Results

## Prediction framework

Phen-Gen combines disease symptoms and sequencing data to estimate the role of genes in rare disorders (Supplementary Fig. 1). The patient symptoms are semantically matched against a database of known human disorders using an in-house implementation of Phenomizer[10]. The potential role of novel but functionally similar genes is estimated using random walk with restart over a gene interaction network. This network was constructed by integrating seven data sources. The sequencing data is analyzed using an exome-centric approach and a genomic approach. The performances of Phen-Gen's predictors for exome and genome have been quantified using enrichment of disease causal mutations, in conjunction with depletion of common polymorphisms. The phenotypic and genotypic predictions are combined and genes ranked based on the two corroboratory pieces of evidence.

The performance of method is first evaluated at the variant level using Human Gene Mutation Database[11] (HGMD 2011.4) reported disease variants across different subclasses (Supplementary Fig. 7). In all three coding categories, 84% (or higher) mutations were predicted damaging with high probability (≥0.9). The genomic predictor was evaluated using HGMD reported disease causal regulatory variants; 80% were scored and 38% were assigned a damaging probability of 0.5 or higher.

## Simulated datasets

We evaluated Phen-Gen with *in silico* patients. For a given disorder, a patient's genetic data was created by adding a HGMD disease causal mutation into a healthy individual's genome or exome. The patient's phenotypic data was generated based on reported disease symptoms. Combining all dominant and recessive simulations, Phen-Gen assigns the disease causal variant first rank in 88% of the cases across different classes of coding variants (Table 1 and Supplementary Table 1).

To ascertain the performance in implicating novel causal genes, we split the HGMD reported variants based on whether the disease-gene link is 'known' or 'unknown' to our local Phenomizer database. When the gene is known, Phen-Gen assigns it the first rank in 92% cases for dominant and 96% cases for recessive diseases (Table 1). However even for unknown disorders, adding the disease symptoms improved the prediction and Phen-Gen is able to correctly identify the true gene in 43% of dominant and 92% of recessive disorders. To further evaluate Phen-Gen's role in novel disease-gene discovery, we masked the gene's association with the disorder in the respective simulation. The results are comparable to the true unknown cases. Combining results from all unknown and masked simulations, Phen-Gen is able to assign first rank to disease causal variants in 71% cases (Supplementary Table 2).

To evaluate any potential biases of using 1000 Genomes for both allele frequency estimation and generating *in silico* patients, the simulations were repeated using only Exome Sequencing Project (ESP)[12] data in the common variant filtration step (Supplementary Table 3). We also evaluated the performance of our method in identifying the disease causal gene with a compound heterozygous disease inheritance pattern (Supplementary Table 4 and Supplementary Fig. 8). Moreover, to account for phenotypic heterogeneity, Phen-Gen was assessed by randomly sampled symptoms from 44 disorders for which the symptom list and their frequency of occurrence has been compiled[10] (Supplementary Table 5 and Supplementary Fig. 9).

It is more challenging to narrow down the role of a particular noncoding variant in a rare disorder. Phen-Gen correctly identifies the genetic cause in 49% of cases with noncoding disease variants, with phenotypic information contributing substantially to the prediction (Supplementary Table 6). Moreover, the decoupled nature of the phenotypic and genotypic predictors within the unifying analysis framework bodes well to improving predictions as the regulatory role of different genomic sites is unraveled.

## Real datasets

We also applied Phen-Gen to a study comprising of a hundred father-mother-child trio families with the child in each family suffering from intellectual disability symptoms[1]. Eleven families with variants implicated in recessive or X-mode inheritance were used in this evaluation. 81% (13/16) of the reported genes were ranked in the top 10 in our prediction (Table 3). In the original study all mutations reported in dbSNP or observed in the in-house dataset were removed from further evaluation. Phen-Gen, on the other hand, allows rare (below 1% MAF) variants from dbSNP. When the original study criteria were adopted, the predictions improved further. 8/11 patients were assigned an implicated gene as the top gene and all reported genes now ranked in the top 5 in the respective families (Supplementary Table 7).

## Comparison

We evaluated Phen-Gen against VAAST[9] and eXtasy[6], both of which rank genes or variants based only on nonsynonymous mutations (Fig. 1). eXtasy under default setting ranks all common and rare variants. In its manuscript it was also evaluated with rare variants (MAF<1%) separately. We show results under both scenarios. VAAST assigns top rank to an average of 14 genes per simulation; whereas Phen-Gen and eXtasy assign a continuum of ranks with a lone gene assigned the top rank in each simulation. Phen-Gen assigns the causal gene top rank in 82% cases for dominant disorders compared to 62% for VAAST and 24% for eXtasy. For recessive disorders, Phen-Gen assigns top rank in 97% cases compared to 78% for VAAST, and 30% for eXtasy.

While Phen-Gen was in submission, PHEVOR[8] was published by the authors of VAAST. PHEVOR combines multiple disease and functional ontologies to improve the performance of genomic predictors like VAAST. To evaluate its performance, 50 dominant and 50 recessive VAAST simulations were randomly chosen and the genes reprioritized using PHEVOR. Phen-Gen identifies the causal gene in 90% of dominant and 88% of recessive simulations, while VAAST+PHEVOR assign the top rank in 66% and 68%, respectively (Fig. 1). In conclusion, Phen-Gen outperforms VAAST, eXtasy, and VAAST+PHEVOR by 19-58%.

PHIVE[7] is another recently published method that integrates phenotypic information. It scores all exonic variants and matches human disease symptoms against a knockout mouse phenotype database using semantic similarity. In its publication, it was evaluated in a similar simulation framework albeit with a different disease set of HGMD variants. The authors report its ability to assign the disease causal gene first rank in 66% of cases for dominant and 83% of cases in recessive disorders. In comparison to PHIVE, Phen-Gen shows 16 and 13 percentage points' advantage in identifying the true causal gene for dominant and recessive disorders respectively.

The impact of regulatory variants in congenital disorders is well established[13]. FunSeq[14] was recently published to predict the regulatory role of noncoding mutations. FunSeq's results using HGMD regulatory variants indicate a strong enrichment of damaging mutations, but only 18% of the damaging regulatory variants fall within its first filtration criteria. In contrast, Phen-Gen assigns damaging probabilities to 80% of these variants (Supplementary Fig. 7).

Phen-Gen highlights the advantage of integrating patient symptom information in understanding disease mechanism and works on both coding and noncoding variation.

# Acknowledgements

The authors would like to thank Radboud University Nijmegen Medical Centre for sharing the 100 intellectual disability patient data in particular J. de Ligt for his help with this data. We would also like to thank S. Köhler for his help with Phenomizer, N. Jinawath for her help interpreting patient symptoms, and S. Prabhakar and N. Clarke for their comments on the genomic predictor. The authors would also like to thank S. Prabhakar, S. Davila, A. Wilm and R. del Rosario for their comments on the manuscript.

# Author Contributions

AJ conceived and designed the project, designed and implemented the analysis framework, implemented methods, conducted experiments, interpreted results, wrote the initial manuscript, revised and proofread the paper. SA implemented methods, conducted experiments, set up the web-server, revised and proofread the paper. PCN conceived and designed the project, revised and proofread the paper, and supervised the project.

# Figure legend

**Figure 1**
**Comparison with VAAST, eXtasy, and VAAST+PHEVOR**
The comparison of Phen-Gen, VAAST, eXtasy and VAAST+PHEVOR in simulations using OMIM disease symptoms and HGMD nonsynonymous mutations is shown. In each panel, the ability of the methods to narrow down the true gene search within 1, 5 and 10 genes is depicted. (a) The two panels reflect comparison with VAAST and eXtasy. For Phen-Gen the bar is split into the predictive power based on genotypic prediction and the added advantage gained from disease symptoms. VAAST only uses the genotype data and assign multiple genes the same rank at the top of the order. For a fair comparison, the true gene was assigned the worst, average and best VAAST rank among similarly ranked peers. For example, if VAAST assigns the true gene along with 4 other genes the same top rank, then the causal gene will be ranked as 1, 3 and 5 for best, average and worst respectively. The three components of the VAAST bar reflect the performance across the three scenarios. eXtasy ranks all rare and common variants. The two components of eXtasy bar reflect its performance under the default settings and the added advantage of discarding common variants (MAF>1%) from its input. (b) 50 dominant and 50 recessive simulations were randomly chosen for comparison with PHEVOR. For VAAST only the best rank (among similarly ranked genes) is depicted. PHEVOR+VAAST bars highlight the added advantage of analyzing VAAST output in light of the phenotypic information by PHEVOR.

**Table 1 Performance in simulated patients for coding predictor**

Phen-Gen's performance is evaluated with OMIM-listed disease symptoms and HGMD- reported variants for coding predictor. Across all simulations Phen-Gen assigns the causal variant top rank in 88% of cases. The results are categorized based on the class of the true causal variant and the table entries reflect the percentage of simulations in each category with the indicated result. The top three rows in each sub-table represent the performance of the complete method (genotype and phenotype), the genotype-only approach, and the phenotype-only approach respectively. The performance of Phen-Gen when the causal gene is previously known and unknown is depicted as well.

|  | Dominant | | | Recessive | | |
| --- | --- | --- | --- | --- | --- | --- |
|  | Top gene (%) | Top 5 genes (%) | Top 10 genes (%) | Top gene (%) | Top 5 genes (%) | Top 10 genes (%) |
| **Missense and Nonsense** | | | | | | |
| Phen-Gen | 82 | 89 | 91 | 97 | 98 | 98 |
| Genotype only | 0 | 3 | 17 | 75 | 98 | 98 |
| Phenotype only | 30 | 66 | 72 | 23 | 49 | 67 |
| Known | 93 | 97 | 98 | 97 | 98 | 98 |
| Unknown | 51 | 67 | 72 | 92 | 97 | 97 |
| **Splice site** | | | | | | |
| Phen-Gen | 80 | 85 | 85 | 87 | 87 | 87 |
| Genotype only | 0 | 4 | 29 | 72 | 87 | 87 |
| Phenotype only | 33 | 75 | 77 | 26 | 50 | 65 |
| Known | 87 | 89 | 89 | 87 | 87 | 87 |
| Unknown | 49 | 67 | 70 | 81 | 82 | 82 |
| **Indels** | | | | | | |
| Phen-Gen | 80 | 88 | 94 | 97 | 98 | 98 |
| Genotype only | 0 | 1 | 2 | 64 | 98 | 98 |
| Phenotype only | 31 | 72 | 74 | 24 | 56 | 72 |
| Known | 94 | 98 | 98 | 97 | 98 | 98 |
| Unknown | 25 | 47 | 75 | 96 | 97 | 97 |
| **Combined (Missense, Nonsense, Splice site and Indels)** | | | | | | |
| Phen-Gen | 81 | 88 | 91 | 96 | 97 | 97 |
| Genotype only | 0 | 3 | 13 | 72 | 97 | 97 |
| Phenotype only | 31 | 69 | 74 | 24 | 51 | 68 |
| Known | 92 | 97 | 97 | 96 | 97 | 97 |
| Unknown | 43 | 61 | 73 | 92 | 96 | 96 |

# Online Methods

## Gene list

We aimed to capture four popular gene lists (Consensus CDS, RefSeq, Ensembl and UCSC known). Incomplete transcripts were discarded in each database and only protein coding genes were considered. Alternatively spliced transcripts reported in the same data source were assigned to the same gene. The mapping from the first three databases to UCSC known gene list was downloaded and utilized to merge gene names across data sources. Additionally, two genes were merged if any of their isoforms exhibited greater than 95% identity in the exon sequence in the same orientation[15]. Finally, an in-house mapping used for identifying alternate isoforms in the SIFT database was employed[16]. The final gene list comprised of 26,803 protein coding genes with 11.4 gene or transcript identifiers reported per gene on average. This gene set was employed in all the experiments in the manuscript and any downloaded gene identifier was translated to this set.

## Variant prediction

For the coding predictor, each called variant in the patient's genome (or exome) is evaluated if it lies within the coding regions in a reported transcript of the fore mentioned gene list, or if it falls within the splice site definition of the intron-exon boundary. The coding variants are further sub-categorized as start-loss, stop-gain, stop-loss, splice site, nonsynonymous, synonymous or indel. Each variant is assigned a probability of deleteriousness based on its highest estimated damaging impact. For the genomic predictor, all coding and noncoding variants are analyzed for their putative functional role.

### Nonsynonymous variants

Nonsynonymous mutations are probably the best studied class of damaging mutations and a plethora of literature has been written and algorithms developed to evaluate their damaging impact. For our prediction we combined two commonly used algorithms SIFT[16] and PolyPhen-2[17]. The aim here is to estimate the probability of a non-synonymous mutation given its' SIFT and PolyPhen-2 scores. We employed a positive and neutral dataset to estimate the empirical distribution of these scores both individually and in conjunction. The positive set comprised of disease causal nonsynonymous mutations from the Human Gene Mutation Database (HGMD)[11]. The neutral set contained nonsynonymous substitutions in the human reference genome with respect to the ancestral sequence as inferred from the human-chimp-gorilla exome alignment. Only the loci where both chimp and gorilla sequence aligned and agreed on the allele were considered. The two distributions were combined using Bayes' rule with prior probability of a nonsynonymous variant being damaging 0.67[18]. The two algorithms exhibit general agreement (Supplementary Fig. 2). Hence the two ends of the spectrum where the predictions disagree are quite sparse. The posterior probability in these bins was interpolated using the geometric mean of their nearest neighbors. PolyPhen-2 predictions tend to average out within each bin; adding SIFT scores helps differentiate and improve the probability estimates within each bin. The continuum of predicted probabilities can be thresholded to achieve 91% accuracy with less than 2.1% false positives; which compares favorably with MutationTaster[19] at 87% accuracy and 4% false positive rate using the same data. It needs to be emphasized though that the downstream analysis uses the probability estimates and not the binary predictions.

*Splice site variants*

Exon-intron boundary information was extracted for all the transcripts in the gene list from their respective databases. Mutations in the 8 base pair locus surrounding the donor site or the 3 base pair locus surrounding the acceptor site (Supplementary Fig. 3) were considered splice disrupting. In particular the 2 base pair at the start and end of an intron tend to be more conserved and hence were treated separately[20]. To estimate the probability of deleteriousness of a mutation affecting splicing boundaries, we employed splice site disease causal mutations from HGMD as a positive set. The neutral set contained common mutations in dbSNP with minor allele frequency greater than 30% within the defined sites. The prior probability of a splice site mutation being damaging was estimated by the reduction in the number of common mutations at these sites in comparison to the whole genome. The rationale being that the mutation rate is assumed to be constant in a random subset of the genome; and this reduction in common mutations is attributed to prior mutations being pruned out by negative selection. The splice site definition fails to capture 8% of mutations reported in HGMD (with another 1% lying outside our gene list). A closer examination of these false negatives reveals that most of these sites are more than 10 bp away from our defined exon-intron boundaries. Some of these could reflect variability in transcript annotation; however others could potentially be introducing alternate splice sites which would not be represented in the prediction.

*Start-loss, Stop-gain and Stop-loss variants*

Start and stop codon altering mutations and nonsense mutations are highly deleterious and in most analysis are assumed to be damaging[21] . We confirmed this hypothesis by computing the relative decrease in common mutations at these sites with respect to the complete genome. Once again, common mutations (MAF>30%) in dbSNP were used and these sites were defined using our gene set. Our analysis reveals that these mutations are highly likely to be damaging with probability 0.999.

*Small insertions and deletions*

To estimate the potential damaging impact of a genic indel, we first classify them as either frameshift or non-frameshift. A different predictor is used for each class. In both cases, the respective disease causal indels from HGMD constitute the positive set. The neutral set comprises of common mutations from dbSNP (MAF>30%). The prior probability of a coding indel being damaging is estimated using 54 unrelated whole genome sequencing samples made publicly available by Complete Genomics Inc. The dataset revealed a strong exonic bias in indel calls, with 20% of indels located within 4% of the genome annotated as coding genes. To rectify for this bias, we compared the enrichment of rare indels (MAF<5%) and common indels (MAF>30%) within and outside the genes. The assumption being that both rare and common indels are impacted equally by the coding bias. Based on this computation, the prior probability of a genic indel being damaging was estimated to be 0.0787. This is a weak prior and the true value is likely to be higher. However because of higher false positive rate in calling indels (in comparison to SNPs) it would be prudent to err on the conservative side. Further the weak prior is rescued by strong observation in both indel categories. All frameshift indels are assigned the probability of deleteriousness based on the empirical distribution of its positive and neutral set combined using Bayes' rule. Non-frameshift indel predictions are further refined by incorporating the importance of the impacted locus (based on its tolerance to single nucleotide mutations).

*Genomic variants*

Next, we aimed to estimate the putative functional role of all genomic variants. To this end, conservation, putative regulatory interactions and proximity to genes were used to annotate the variants (Supplementary Fig. 4). Both GERP++[22] and PhyloP[23] (threshold 1.445) were used to define evolutionary conserved sites. ENCODE computational predictions for transcription factor binding sites and experimental annotations for DNASE hypersensitive sites were included. Proximity to gene annotations (including coding sequence, UTRs and 70 bp at the start and end of each intron) was incorporated.

A key statistical challenge in estimating the functional role of noncoding mutations is the dearth of disease causal regulatory mutations in public datasets. Further, most studies focus on coding variants and hence dbSNP has accumulated a higher number of these variants over time. To correct for both these shortcomings, we employed two positive and two neutral datasets. Disease causal regulatory mutations in HGMD and reported GWAS hits (downloaded from NCBI) constituted the positive sets. The neutral sets contained common SNPs in dbSNP and common SNPs in publicly available unrelated sample data from Complete Genomics Inc (MAF > 0.30) respectively. Ten thousand permutations were performed and each positive and neutral dataset was randomly selected with equal probability. Each selected dataset was subsampled, in each permutation, to estimate confidence intervals (Supplementary Fig. 5). The positive and neutral empirical distributions thus estimated were combined within Bayes' rule. The prior probability of a random mutation in our annotated regions being damaging was estimated to be 0.0688. In addition to the fore mentioned annotations, we also considered further annotations including GUMBY[24], noncoding RNA from GENCODE[25], microRNA[26] and GC content[27]. The technical difficulty in embracing more annotations stems from the small size of the positive sets. Adding more annotations increases the number of bins exponentially. This dilutes the signal within each bin, making it more susceptible to stochastic noise. GUMBY is stringent in its conservation threshold and thus quite accurate but not all-encompassing to all functional regions. Furthermore, 99.65% of GUMBY conserved sites overlapped with GERP++. MicroRNA predicted loci[28] and GENCODE annotations[29], despite their well established epigenetic role were not well represented in our positive sets. GC content was suggestive of the regulatory role when only HGMD was used as positive set; the predictive power disappeared when both positive sets were employed. This could potentially reflect the enrichment of de novo mutations in the HGMD dataset, as mutation rate across the genome tends to correlate with GC content[27]. These predictors can be further interrogated and perhaps rescued with increasing size of regulatory variant datasets.

## Pooling

The estimated loss of function at the genic level is predicted by pooling damaging variants within each gene considering the most damaging predicted variants for maternally and paternally inherited chromosomes. If the haplotype phase information is not available, the first and second highest variants are considered under the assumption that they lie on opposite chromosomes. Noncoding mutations are assigned to the nearest gene within 50 kb. This criterion is similar to conservative settings in GREAT[30] and assumes potential cis regulatory impact of these variants on the respective genes. The inheritance pattern of the variants is compared against the disease inheritance pattern. For small pedigrees (such as trios or quartets) only variants consistent with this pattern are evaluated. For larger pedigrees, the flexibility to allow for nominal inconsistencies is permissible. This leeway is to account for potentially undiagnosed patients, incomplete penetrance, low coverage, and possible error in variant calls. Common variants are omitted from this pooling. For the prediction, any variant with reported minor allele frequency above 1% in either 1000 genomes[31], National Heart, Lung, and Blood Institute Exome Sequencing Project (NHLBI ESP) or dbSNP[32] version 135 is deemed common and excluded from consideration as the etiological variant. The analysis allows for the user to discard or incorporate variants inconsistent with the pedigree. Unless de novo mutation is the likely cause of the condition, the former is recommended.

Healthy individuals have been reported to harbor up to 100 damaging mutations[33]. The distribution of these so-called deleterious variants is not uniform across all genes and certain genes (such as olfactory genes) are more likely to carry these loss-of-function variants. To account and rectify for these incidental red herrings, we compute a null distribution of our predictor for each gene using the samples from 1000 Genomes. These individuals have not been reported to suffer from any disorders and represent a generic snapshot of human genetic variability across the globe. The pooled probability distribution of each gene is estimated under both the dominant and recessive inheritance pattern (Supplementary Fig. 6). For a patient, only genes harboring variants which exceed the ninety ninth percentile of the corresponding null distribution are considered for downstream analysis. For example, more than 1% of 1000 Genomes samples carry rare variants in *CYP2C19* on both chromosomes with estimated damaging probability 0.88 or higher. Hence for a recessive disorder patient, *CYP2C19* will only be considered as a candidate gene if there are two variants within the gene that both have a damaging probability higher than 0.88. A key advantage of our approach, in the context of the method, is that we establish a continuum of predicted loss of functionality of each gene within healthy individuals.

## Phenotypic prediction

The patient's disease symptoms are mapped to Human Phenotype Ontology database using an in-house replica of Phenomizer[10]. The aim here is to match the patient's symptomatology to the list of known disorders and estimate the significance of each disease match. The Benjamini-Hochberg multiple testing corrected p-values are translated to probabilities assuming disease set has a uniform prior.

$$P_{\text{disease}}(i) = \frac{\frac{1}{\text{\# of disorders}} \times P(\text{disease match}|\text{true match})}{\left(\begin{array}{c}\frac{1}{\text{\# of disorders}} \times P(\text{disease match}|\text{true match}) \\ + \frac{(\text{\# of disorders} - 1)}{\text{\# of disorders}} \times P(\text{disease match}|\text{false match})\end{array}\right)}$$

$$P_{\text{disease}}(i) \approx \frac{\frac{1}{\text{\# of disorders}}}{\frac{1}{\text{\# of disorders}} + P_{\text{value}}(i)}$$

Each disease probability is assigned to all genes implicated in literature for that disorder. If no genes are currently known, the probability is distributed uniformly across all genes. Certain genes are known to impact multiple disorders. Pleiotropy is accounted for by combining all the assigned probabilities for a gene across the spectrum of disorders it has been reportedly involved in.

$$P_{\text{gene}}(j) = \left(1 - P_{\text{disease}}(i)\right) \times \delta(i,j)$$

$$\text{where } \delta(i,j) = \begin{cases} 1 & \text{if gene}(j) \text{ has been implicated for disease } i \\ 0 & \text{otherwise} \end{cases}$$

Thus far we have estimated the role of each gene leveraging on direct knowledge of their involvement in diseases with similar symptoms. Next we incorporate 'guilt by association' based on evidence of their interaction with a known perpetrator. To gather this circumstantial evidence, a random walk with restart is conducted on gene-gene interaction network. Each gene is selected as a starting point of the walk with probability $P_{\text{gene}}$. Restart probability of 90% is used for the walk. This translates to 90% of the phenotypically matched probability aka 'guilt' retained by the initial gene and 10% of it permeated across its known associates; with stronger evidence of interaction leading to a higher probabilities.

## Gene-gene interaction network construction

In the context of the method, the gene-gene interactions are not just limited to physical interaction of genes. Rather they reflect the ability of two genes to impact the same underlying biology and thus lead to the same (or similar) disorders and symptoms. This information is agglomerated across different sources including known involvement in the same pathway or ontology domains, data-mined reports of co-occurrence across publications, and/or correlated co-expression among tissue types. The gene-gene interaction network is constructed by combining these sources in a framework inspired by ref. 34. A key difference is that the original study made binary predictions, thus representing an un-weighted graph; whereas we construct a randomized graph with edge weights proportional to confidence in interaction. The network construction is as follows. All pairs of genes reportedly involved in the same pathway (REACTOME[35], KEGG[36], NCI-Nature[37]) are considered true interactions. This confidence stems from the manual curation process of these sources by domain experts. To improve recall of the network, we supplement the pathway information with additional data sources. The pathway based interaction are used to evaluate the predictive power of data mining approaches (BioGRID[38], STRING[39]), Gene Ontology domains annotations[40] (cellular component, molecular function, and biological process) and gene co-expression (COXPRESdb[41]). These predictions are combined within a unifying framework in a manner similar to ref. 34 with prior probability of two random genes interacting estimated from pathway databases. A high confidence (co-involvement probability >=0.1) gene-gene interaction network was thus defined. In total, the network comprised of 920,898 interactions.

To test the efficacy of this network in representing genetic heterogeneity, gene pairs implicated for the same disorder were extracted from HGMD and OMIM (10,852 and 5,083 respectively). 31% and 39% of these pairs are represented by a direct edge in the network ($P<<10^{-16}$ based on distribution of random gene pairs) indicating that the probability of their involvement in similar diseases is well represented in the network; and that the network is significantly enriched in true interactions; which bodes well in its efficacy to elucidate novel gene involvements in known disorders.

Disease causal genes tend to play a more central role in interactome. This is reflected in the gene-gene interaction network where HGMD and OMIM implicated genes tend to have a higher number of edges ($P \leq 10^{-166}$ and $P \leq 1.8 \times 10^{-192}$ respectively).

## Combined

Assuming phenotypic and genotypic probability estimates are independent observations of the same underlying probability of involvement of a gene we combine the two predictions for each gene within a Bayesian framework.

$$Probability\ gene(j)\ is\ disease\ causal = \frac{P_G(j) \times P_P(j)}{P_G(j) \times P_P(j) + (1 - P_G(j)) \times (1 - P_P(j))}$$

where $P_G(j)$ = probability gene($j$) is disease causal based on genotype data
$P_P(j)$ = probability gene($j$) is disease causal based on phenotype data

## Computational efficiency

Exome sequencing routinely yields 20,000 variants per sample; the number increases to 4 million for whole genome sequencing. We employ a simultaneous linear scan of the variant and annotation files. To speed up the process 'regions of interest' were defined using interval forests; with one tree per chromosome. Overlapping intervals were merged to define a sieve which represents 19% of the genome. This allows the implementation to quickly sift out the variants of interest discarding the rest. To expedite the actual run-time execution, whenever possible, the burden of computation was moved to pre-processing and information stored in a binary format for faster reading, e.g. for noncoding variants the database overlap information is encoded in 1 byte per interval.

The source code, preprocessed databases and executables are provided as single downloadable package. This includes an in-house implementation of Phenomizer with limited functionality necessary for Phen-Gen. The source code is released under the GNU General Public License, so a computationally savvy user can make changes to incorporate further metrics of deleteriousness and improve on the model. We plan to periodically update the online and downloadable versions of Phen-Gen with each major release of the source databases. The software is also available as an online server for ease of use.

## Simulating *in silico* patients

To evaluate Phen-Gen we generated *in silico* patients. For a given disorder, a patient's genetic data was created by adding a known disease causal mutation from HGMD into a healthy individual with zygosity consistent with the reported disease inheritance pattern. These genomes on average harbor four million genomic variants, of which twenty four thousand are coding[42]. The patient's phenotypic data was generated based on reported disease symptoms. Phen-Gen allows pedigree data, which considerably narrows down the search pool but in the simulations we aimed for the extreme case where only the lone patient information is available. In the first set of simulations, 765 diseases for which at least one causal variant was reported in HGMD and the disease symptoms were defined in OMIM[4] were extracted. These symptoms were used to define the patient's condition. Each causal variant was systematically spiked in each of the 1092 individuals from 1000 Genomes dataset in turn and the remaining individuals were used to define the null distribution (Tables 1 and Supplementary Tables 1 and 6).

To evaluate Phen-Gen's performance in identifying the disease causal gene with a compound heterozygous disease inheritance pattern a second set of simulations was conducted. In these, the general framework remained the same as before. But this time instead of a single variant, two variants were added to the patient genome in a heterozygous state on opposite chromosomes. For these simulations only recessive inheritance pattern disorders with at least two variants reported in (or near) the same gene were used with a simulated patient generated for every pair of reported variants (Supplementary Table 4). In the case where one or both of the members of a spiked variant pair was noncoding, the genomic variant predictor was used (Supplementary Fig. 8).

To evaluate Phen-Gen's robustness to phenotypic heterogeneity, 44 disorders for which the symptom list and their frequency of occurrence has been compiled[11] were used. For each disorder, five medical histories were constructed by randomly sampling each symptom based on its frequency. The simulations were conducted in a similar manner as before, with the added caveat that every individual patient genome was evaluated five times, once with each medical history (Supplementary Table 5).

# Methods-only references

35. Matthews, L. *et al. Nucleic Acids Res.* **37,** D619–D622 (2009).

36. Kanehisa, M., Goto, S., Sato, Y., Furumichi, M. & Tanabe, M. *Nucleic Acids Res.* **40,** D109–14 (2012).

37. Schaefer, C.F. *et al. Nucleic Acids Res.* **37,** D674–D679 (2009).

38. Stark, C. *et al. Nucleic Acids Res.* **34,** D535–D539 (2006).

39. Franceschini, A. *et al. Nucleic Acids Res.* **41,** D808–15 (2013).

40. Ashburner, M. *et al.* The Gene Ontology Consortium. *Nat. Genet.* **25,** 25–29 (2000).

41. Obayashi, T. *et al. Nucleic Acids Res.* **41,** D1014–20 (2013).

42. Altshuler, D.M. *et al. Nature* **491,** 56–65 (2012).
21

# Competing Financial Interests

The authors declare no competing financial interests.

# Phen-Gen: Combining Phenotype and Genotype to Analyze Rare Disorders

Asif Javed[1], Saloni Agrawal[1], Pauline C. Ng[1]

[1]Computational & Systems Biology, Genome Institute of Singapore, Agency for Science, Technology and Research, Singapore

# Supplementary material

## VAAST, PHEVOR and eXtasy simulations

The performance of Phen-Gen was compared against VAAST[1], PHEVOR[2], and eXtasy[3]. PHEVOR is a phenotype-based add-on tool which relies on genomic predictions from methods like VAAST. Both VAAST and eXtasy predict only on the damaging impact of amino acid changing SNVs and do not score indels, splice or noncoding variants. Hence for this comparison only nonsynonymous mutations in Human Gene Mutation Database (HGMD) were used. VAAST uses a unique file format (Genome Variant Format, or GVF) and their analysis package comes with conversion software to convert from VCFs to this format. Unfortunately, we faced compatibility issues with VCFs extracted for majority of the samples from African American population (ASW), within the 1000 Genomes data. To keep things simple, all ASW samples were removed from these. For computational efficiency, a random set of 1000 individuals were set aside as controls and 26 individuals were used to generate the simulated patients. For each causal variant, one individual was randomly selected and eXtasy, VAAST and Phen-Gen were run independently. The comparison was conducted using HGMD reported disease causal variants for which the disease symptoms are defined in OMIM.

For VAAST the 1000 individuals served as controls to estimate the composite likelihood under the healthy model to be compared against the disease model. Phen-Gen and eXtasy provide a continuum of rankings whereas VAAST (on average) identifies 14 genes at the top of the list. For a fair comparison, the true causal gene for VAAST ranking was assigned best, average, and worst rank among similarly ranked genes. For example, if VAAST assigns the true gene along with 4 other genes the same top rank. The causal gene will be ranked as 1, 3 and 5 for best, average and worst respectively. This is reflected in the three components of the bar in VAAST results (Fig. 1). VAAST was able correctly narrow down the true causal gene to within 14 genes (on average) in 62% of dominant and 78% of recessive cases. In comparison, Phen-Gen was able to identify the correct etiological gene in 82% cases for dominant and in 97% cases for recessive OMIM reported disorders.

PHEVOR was recently published by the authors of VAAST. It combines information from multiple disease and functional ontology databases to re-prioritize genomic predictions in light of patient-specific symptoms. PHEVOR is only available as an online tool; making a comprehensive simulation comparison not possible. In its publication, PHEVOR was evaluated in 100 simulations. We chose the same number for our comparison. 50 dominant and 50 recessive cases were randomly chosen from the VAAST simulations. As a sanity check, VAAST predictions of this subset were compared against VAAST results for all simulations to ensure that the subsample reflects the general trend. PHEVOR web server only allows up to 5 disease symptoms. If the phenotype description exceeds this number, 5 symptoms were randomly chosen. The combination of VAAST and PHEVOR was able to assign the causal gene top rank in 66% of dominant and 68% of recessive simulations. In comparison for the same cases, Phen-Gen was able to correctly identify the causal gene in 90% of dominant and 88% of recessive simulations (Fig. 1).

The better performance of Phen-Gen can be attributed to methodological differences as well as difference in data sources integrated in the prediction. PHEVOR's 'ontology propagation' does not take in account the information content of its various data sources. Phen-Gen estimates this in the gene interaction network construction by relying on pathway databases. Pathway databases are often the best curated gene functional information available[4]. PHEVOR currently does not use this resource, although its authors concede integrating pathway information to be an active area of PHEVOR's development.

eXtasy relies on locus and gene specific information and does not use a control population. Hence the control set was not used in its prediction. Under default settings it evaluates all variants both common and rare. In its manuscript it was evaluated with rare variants (MAF<1%) as well. eXtasy's performance is thus evaluated under both scenarios. In the first, all variants in the individual exome were provided as an input. In the second, only rare variants were used and variants with MAF>1% in 1000 Genomes, dbSNP or National Heart, Lung, and Blood Institute's Exome Sequencing Project (ESP) were discarded. eXtasy is able to narrow down the causal variant within top 10 variants in 69% of the cases for dominant and 76% of the cases for recessive disorders. In comparison Phen-Gen is able to narrow down the causal variant within 10 variants in 91% of the cases for dominant and 98% of the cases for recessive disorders. The relative advantage of Phen-Gen can potentially stem from multiple factors. eXtasy does not incorporate the disease inheritance pattern in its prediction. It also ranks variants based on individual phenotypes. Although the information is combined in rank aggregation, the interplay of different combination of phenotypes may not be well represented.

For the comparison Phen-Gen used the identical control set of 1000 individuals from VAAST to define the null distribution of genes. Phen-Gen outperforms eXtasy, VAAST and VAAST+PHEVOR by 19-58% (Fig. 1). eXtasy currently only allows a subset of disease symptoms defined in Human Phenotype Ontology. To investigate Phen-Gen's relative advantage due to a more comprehensive symptom list, Phen-Gen simulations were repeated restricting it to the symptoms accepted by eXtasy; Phen-Gen's ability to identify the true disease causal gene dropped by 1% in these simulations (results not shown).

For further comparison with VAAST, 44 phenotypic heterogeneous dataset was employed. The results show the relative performance and highlights Phen-Gen's advantage even when disease symptoms are not completely specific (Supplementary Fig. 8).

## Performance using real dataset

Phen-Gen was evaluated using a recently published real dataset comprising of one hundred parents-child trio families with the children exhibiting intellectual disability symptoms[5]. This data contains variant calls for each family, as well as a rich resource of medical history detailing each patient's unique symptoms. The information was initially electronically parsed, and then manually evaluated by two of the authors independently to absolve the translation to Human Phenotype Ontology terms from any ambiguities. Further advice from a clinician was sought to remove any errors of interpretation.

On the genotypic side, a key challenge in handling the real dataset was the noise level in the variant calls. In the original study itself, the focus was on de novo mutations and the provided 'high quality calls' cast an initial wide net to encompass large number of candidate mutations. In the original publication, these calls were pruned by further bioinformatic processing and finally bench validation. Replicating this effort required access to the patient samples themselves, which was not possible. The validation rate of de novo mutations reported even after further bioinformatic processing was 11.4%[3]. To sidestep validation issues due to noise level in the data and focus on the downstream analysis that Phen-Gen aims to provide, we initially focused on eighteen families with variants implicated in recessive or X-mode inheritance. Seven families were further removed from consideration as the variants reported in the publication were not observed in the correct inheritance pattern in the provided data and had likely been corrected in the bench validation[3]. Hence eleven families were used in this current study.

The performance of Phen-Gen is evaluated for variants reported in the original publication (Supplementary Table 7). Phen-Gen allows for variants with MAF below 1% in public databases. The first column of Phen-Gen results depicts the performance in this scenario. The original study however discarded all known variants (already existing in dbSNP, or observed in their in-house database). After adopting the same filtering criteria as the original study, 16/21 genes agreed with the original study, and only five are potentially false positives. Of these five variants, two were homozygous in the patient and observed in heterozygous state in two different families and the remaining three are only observed in the respective family. Of the three novel variants two are indels which were not evaluated in the original study. These five variants cannot be ruled out as disease causal candidates based purely on the bioinformatics analysis of the variant calls and further bench validation and functional analysis would be needed to implicate or exonerate them.

Intellectual disability is a genetically diverse disorder and it has been estimated that more than a thousand different genes may be playing a role[6]. Thus despite the rich and detailed patient history, it is challenging to list it down to a few genes based purely on symptomatic knowledge. We analyzed the symptoms of 58 patients with reported recessively inherited deleterious mutations, de novo mutations, or X-linked mutations in males (Supplementary Tables 3 and 9 in the original publication[5]). The analysis relied only on just the phenotypic information to rank the genes for each patient. To quantify the performance of our approach, we computed the sum of the ranks of all reported genes in the respective patient's phenotypic match. The aim of this analysis is to highlight that the genes reported in the original study based on genetic evidence tend to have a significantly lower rank in our lists than observed just by chance. Each individual patient's rank should have a uniform distribution under the null hypothesis and hence their sum - an Irwin-Hall distribution- is approximated by a Normal distribution. The results indicate that the rank sum of reported genes is significantly lower than expected by chance ($P \leq 0.0036$).

The authors split the gene set into known, unknown and candidate genes. The 'known genes' had been implicated for intellectual disability in prior literature. 'Candidate genes' had not been directly reported for intellectual disability, but are linked to brain and embryonic development and there is further evidence of their involvement. The remaining genes are deemed 'unknown'. Even when we focused on individual gene subclasses, the rank sums in all three categories were significantly lower: known genes ($P \leq 2.44 \times 10^{-8}$), candidate genes ($P \leq 0.0051$), and unknown genes ($P \leq 0.0192$). In the original study, only known genes carrying deleterious variants were considered confirmed as diagnosed; or if the same candidate gene harboring damaging mutations was observed in two different patients with similar symptoms. Known genes implicated in these patients are expected to be ranked lower. Our results indicate that although the unknown genes set could harbor a lot of incidental genes, they are likely to contain some true positives as well; and the diagnostic yield of this dataset can potentially be improved by further corroboration.

## Comparison with PHIVE

PHIVE[7] was evaluated in a simulation framework similar to the one presented in this paper. The authors used a somewhat different HGMD version (869 diseases versus 765 for Phen-Gen) and this could potentially contribute to the difference in results. Since PHIVE is only available as an online server and not a downloadable package, it is not possible to do a comprehensive evaluation using the same dataset. The authors report 66% power to identify the causal gene in dominant and 83% in recessive disorders. Comparison of these reported results reveals Phen-Gen has 13-16% higher efficacy in identifying the true causal mutation in both dominant and recessive disorders (Table 1). This advantage may be attributed to different factors including the underlying prediction framework and better representation of the disease symptoms in human databases.

PHIVE uses phenotypes from mouse gene knockout experiments to establish phenotype to gene links. Using the mouse model to represent human phenotypes has a few limitations. Mouse knockout experiments have only been conducted for approximately a third of the human genes. There has been increasing focus in the recent past to generate a more comprehensive resource[8], and this would improve PHIVE predictions over time. In comparison human disease-gene association span only about 10% of the genes. Both these numbers support the usage of gene (or protein) interaction networks to extrapolate predictions to the remaining set of genes. PHIVE assigns a uniform phenotypic score to the 2/3$^{rd}$ gene set and the authors concede that integrating protein interaction information is a potential future direction to improve results. A second issue which would be more difficult to address is that human and mice do not share disease morphology for all disorders[9]. In particular it would not be representative of any primate specific, or even more constrained human-specific, traits. Ward and Kellis recently showed that a large number of human regulatory regions do not show evolutionary conservation in primate evolution[10], so the mouse model may not recapitulate recently acquired regulatory evolving traits. Despite these limitations, animal models provide an invaluable resource and integrating this information along with human disease matches is one avenue to improve Phen-Gen's predictions.

On the genotypic side, PHIVE assigns arbitrary pathogenicity scores for the different classes of coding mutations (all except missense). These scores were chosen to give optimal predictions within their simulation framework. Their performance declined by almost half in simulations spiking the causal variant in in-house exomes (Fig. 4a of the publication[7]). The authors attributed this decline to their reliance on allele frequency information which would be unavailable for novel variants observed in the in-house data. This issue will be faced during the analysis of any new dataset. Another factor potentially playing a role is that the 1000 Genomes data is highly curated and fine-tuning the predictions based on this dataset may have led to overfitting. For example, the indel calls in the public data have been improved in extensive bench validations[11], whereas in a practical scenario these calls tend to be more error prone.

## Comparison with FunSeq

FunSeq[12] evaluates the regulatory role of a noncoding mutation using a combination of functionality categories similar to Phen-Gen. It identifies the top 0.4% of the genome as 'sensitive' to regulatory disrupting mutations. The selective constraint in each combination of annotations is estimated using enrichment of rare variants. FunSeq annotations are highly predictive of functionality but may not necessarily be all-encompassing. The method follows multiple screening steps to reduce the number of candidate mutations. Analysis of HGMD reported disease causal regulatory variants reveals that their initial filter (representing coding and sensitive regions) only captures 18% of these damaging variants and would annotate the rest as benign. Phen-Gen on the other hand assigns a continuum of probabilities to 80% of the variants. These variants are assigned a lower probability based on genotypic information (Supplementary Fig. 7). These variants would likely be poorly predicted by any genotype only approach and would require phenotypic support for improved predictions. Simulation results support this hypothesis; Phen-Gen is able to specify the true causal variant regulatory mutation in 49% of the cases with phenotype information (Supplementary Table 6). This predictive power stems from phenotypic corroboration with a 30 percentage points advantage in predicting the true causal gene compared to a purely genotype only approach; thus highlighting Phen-Gen's advantage of integrating this viable resource. In 69% of the cases, the disease gene appeared in Phen-Gen's top 10 list of candidate.

## Prior probability for genomic predictor

It is estimated that about 10-15% of the genome is functionally active[13]. Assuming the average of these estimates and that all functional loci reside within our regions of interest span 19% of the genome, we can compute the probability of a random annotated locus being functional to be 12.5/19 = 0.66. Next to compute the probability of a mutation at a functional region being deleterious, we used the PhyloP conserved bases as surrogates of the functional genome and computed the reduction in common mutations in dbSNPs (MAF>30%) at these sites in comparison to the rest of the genome. These numbers were combined to yield the probability of an annotated variant being damaging 0.0688 which is used as prior in the genomic predictor.

## Incorporating pedigree information

For a dominant disease inheritance pattern genes harboring one or more damaging mutations are evaluated. Predicted damaging variants with one or more copies of the alternate allele in all cases and no copies of alternate allele in controls are considered. Only genes with a predicted probability higher (or equal) in cases than controls are incorporated in the downstream analysis.

For a recessive or compound heterozygous disease inheritance pattern, genes harboring two or more damaging mutations are evaluated. Predicted damaging variants with one or more copies of the alternate allele in all cases and one or zero copies of the alternate allele in controls are considered. Only genes with a predicted probability higher in cases than controls are incorporated in the downstream analysis. This allows for compound heterozygosity while reducing the false positives. If both parents are included in the analysis, there is a further filter to require at least one variant from each parent.

Phen-Gen allows the user to restrict the analysis to variants consistent with the pedigree structure of the family. This is the recommended settings as pedigree inconsistent variants harboring potential de novo mutations also tend to be enriched in sequencing errors. The software allows the user with leniency in this criterion by allowing pedigree-inconsistent variants if de novo mutations are the likely cause of the condition. In practice for real datasets however it is highly recommended to use a pedigree based variant caller (such as GATK 2 PhaseByTransmission walker) to prune out false de novo calls.

## Evaluating null distribution of genes

To evaluate Phen-Gen's null distribution of genes we compared our predicted deleterious variant harboring genes with loss of function genes reported in ref. 11 and Residual Variation Intolerance Score (RVIS) reported in ref 14. Ref. 11 used a subset of 1000 Genomes individuals for their analysis and the predicted damaging variants were subsequently curated after further bench validations. Ref. 14 used NHLBI ESP allele frequencies and corrected for gene size as larger genes are more likely to harbor incidental so-called damaging variants due to size. A cutoff of ninety fifth percentile was employed to their set of damaging variant harboring genes. Phen-Gen employs a one percentile cutoff for the null distribution. Despite methodological and dataset differences, genes which exceed this cutoff showed high enrichment in the respective datasets ($P \leq 8\times10^{-3}$ for McArthur et al and $10^{-2}$ for Petrovski et al using Fisher's exact test).

**Supplementary Table 1**
**Additional statistics for results reported in Table 1**

Performance in simulated patients with OMIM-listed disease symptoms and HGMD reported variants for coding predictor. The table is an extension of Table 1 with performance for genotype and phenotype-only predictors added for both known and unknown diseases in each category (see highlighted rows). The number of variants in each category is also included at the top of each table. The performance of both genotypic and phenotypic predictors in each category is presented. The percentage of known (gene disease association in the local Phenomizer database) and unknown variants in each category is also reported.

|  |  | Dominant | | | | Recessive | | |
|---|---|---|---|---|---|---|---|---|
|  |  | Top gene (%) | Top 5 genes (%) | Top 10 genes (%) |  | Top gene (%) | Top 5 genes (%) | Top 10 genes (%) |
| **Missense & Nonsense** |  | 9,194 variants, 1092 individuals | | | | 11,028 variants, 1092 individuals | | |
| Phen-Gen | All | 82 | 89 | 91 | All | 97 | 98 | 98 |
| Genotype only | | 0 | 3 | 17 | | 75 | 98 | 98 |
| Phenotype only | | 30 | 66 | 72 | | 23 | 49 | 67 |
| Phen-Gen | Known (74%) | 93 | 97 | 98 | Known (91%) | 97 | 98 | 98 |
| Genotype only | | 0 | 3 | 19 | | 75 | 98 | 98 |
| Phenotype only | | 37 | 80 | 84 | | 25 | 51 | 68 |
| Phen-Gen | Unknown (26%) | 51 | 67 | 72 | Unknown (9%) | 92 | 97 | 97 |
| Genotype only | | 0 | 3 | 13 | | 72 | 96 | 97 |
| Phenotype only | | 12 | 26 | 40 | | 13 | 29 | 54 |

| **Splice site** |  | 1,581 variants, 1092 individuals | | | | 1,899 variants, 1092 individuals | | |
|---|---|---|---|---|---|---|---|---|
| Phen-Gen | All | 80 | 85 | 85 | All | 87 | 87 | 87 |
| Genotype only | | 0 | 4 | 29 | | 72 | 87 | 87 |
| Phenotype only | | 33 | 75 | 77 | | 26 | 50 | 65 |

| | | Dominant | | | | Recessive | | |
|---|---|---|---|---|---|---|---|---|
| | | Top gene (%) | Top 5 genes (%) | Top 10 genes (%) | | Top gene (%) | Top 5 genes (%) | Top 10 genes (%) |
| Phen-Gen | Known (81%) | 87 | 89 | 89 | Known (94%) | 87 | 87 | 87 |
| Genotype only | | 0 | 4 | 30 | | 72 | 87 | 87 |
| Phenotype only | | 39 | 87 | 89 | | 28 | 52 | 67 |
| Phen-Gen | Unknown (19%) | 49 | 67 | 70 | Unknown (6%) | 81 | 82 | 82 |
| Genotype only | | 0 | 4 | 25 | | 72 | 82 | 82 |
| Phenotype only | | 5 | 20 | 26 | | 3 | 16 | 43 |

| Indel | | 5,972 variants, 1092 individuals | | | | 4,220 variants, 1092 individuals | | |
|---|---|---|---|---|---|---|---|---|
| Phen-Gen | All | 80 | 88 | 94 | All | 97 | 98 | 98 |
| Genotype only | | 0 | 1 | 2 | | 64 | 98 | 98 |
| Phenotype only | | 31 | 72 | 74 | | 24 | 56 | 72 |
| Phen-Gen | Known (80%) | 94 | 98 | 98 | Known (91%) | 97 | 98 | 98 |
| Genotype only | | 0 | 1 | 1 | | 64 | 98 | 98 |
| Phenotype only | | 38 | 87 | 89 | | 26 | 59 | 73 |
| Phen-Gen | Unknown (20%) | 43 | 61 | 73 | Unkown (9%) | 92 | 96 | 96 |
| Genotype only | | 0 | 2 | 11 | | 70 | 95 | 96 |
| Phenotype only | | 8 | 21 | 32 | | 11 | 27 | 56 |

| Combined | | 16747 variants, 1092 individuals | | | | 17147 variants, 1092 indviduals | | |
|---|---|---|---|---|---|---|---|---|
| Phen-Gen | All | 81 | 88 | 91 | All | 96 | 97 | 97 |
| Genotype only | | 0 | 3 | 13 | | 72 | 97 | 97 |

|  |  | Dominant | | | | Recessive | | |
| --- | --- | --- | --- | --- | --- | --- | --- | --- |
|  |  | Top gene (%) | Top 5 genes (%) | Top 10 genes (%) |  | Top gene (%) | Top 5 genes (%) | Top 10 genes (%) |
| Phenotype only |  | 31 | 69 | 74 |  | 24 | 51 | 68 |
| Phen-Gen | Known (77%) | 92 | 97 | 97 | Known (91%) | 96 | 97 | 97 |
| Genotype only | | 0 | 3 | 14 | | 72 | 97 | 97 |
| Phenotype only | | 38 | 83 | 86 | | 26 | 53 | 69 |
| Phen-Gen | Unknown (23%) | 43 | 61 | 73 | Unknown (9%) | 92 | 96 | 96 |
| Genotype only | | 0 | 2 | 11 | | 70 | 95 | 96 |
| Phenotype only | | 8 | 21 | 32 | | 11 | 27 | 56 |

# Supplementary Table 2
# Performance in novel disease gene discovery

The simulations are conducted with OMIM-listed disease symptoms and HGMD reported variants. In each category, Phen-Gen is evaluated with the knowledge of the respective known disease gene association masked from the simulation. The results are highlighted. Across all masked and unknown simulations Phen-Gen assigns the causal variant top rank in 71% of cases. For a comparison, Phen-Gen's performance in the known and unknown categories is also included from Supplementary Table 1. The results indicate a drop in performance for novel gene discovery in comparison to known associations, and highlight comparable performance to the true unknown cases. Since prior disease knowledge impacts only the phenotypic part of the prediction, Phen-Gen's prediction performance based solely on the phenotype is also included.

|  |  | Dominant | | | Recessive | | |
|---|---|---|---|---|---|---|---|
|  |  | Top gene (%) | Top 5 genes (%) | Top 10 genes (%) | Top gene (%) | Top 5 genes (%) | Top 10 genes (%) |
| **Missense & Nonsense** | | | | | | | |
| Known | Phen-Gen | 93 | 97 | 98 | 97 | 98 | 98 |
| Known | Phenotype only | 30 | 66 | 72 | 25 | 51 | 68 |
| Masked | Phen-Gen | 58 | 78 | 84 | 90 | 97 | 98 |
| Masked | Phenotype only | 13 | 23 | 26 | 12 | 17 | 18 |
| Unknown | Phen-Gen | 51 | 67 | 72 | 92 | 97 | 97 |
| Unknown | Phenotype only | 12 | 26 | 40 | 13 | 29 | 54 |
| **Splice site** | | | | | | | |
| Known | Phen-Gen | 87 | 89 | 89 | 87 | 87 | 87 |
| Known | Phenotype only | 39 | 87 | 89 | 26 | 50 | 65 |
| Masked | Phen-Gen | 47 | 61 | 64 | 83 | 87 | 87 |
| Masked | Phenotype only | 5 | 18 | 20 | 14 | 22 | 23 |
| Unknown | Phen-Gen | 49 | 67 | 70 | 81 | 82 | 82 |
| Unknown | Phenotype only | 5 | 20 | 26 | 3 | 16 | 43 |

|  |  | Dominant | | | Recessive | | |
|---|---|---|---|---|---|---|---|
|  |  | Top gene (%) | Top 5 genes (%) | Top 10 genes (%) | Top gene (%) | Top 5 genes (%) | Top 10 genes (%) |
| **Indel** | | | | | | | |
| Known | Phen-Gen | 94 | 98 | 98 | 97 | 98 | 98 |
| Known | Phenotype only | 38 | 87 | 89 | 26 | 59 | 73 |
| Masked | Phen-Gen | 55 | 69 | 74 | 88 | 97 | 97 |
| Masked | Phenotype only | 12 | 23 | 25 | 11 | 21 | 23 |
| Unknown | Phen-Gen | 25 | 47 | 75 | 96 | 97 | 97 |
| Unknown | Phenotype only | 2 | 12 | 16 | 7 | 25 | 65 |

|  |  | Dominant | | | Recessive | | |
|---|---|---|---|---|---|---|---|
| **Combined** | | | | | | | |
| Known | Phen-Gen | 93 | 97 | 97 | 96 | 97 | 97 |
| Known | Phenotype only | 34 | 76 | 80 | 25 | 53 | 69 |
| Masked | Phen-Gen | 56 | 73 | 78 | 89 | 96 | 97 |
| Masked | Phenotype only | 12 | 23 | 25 | 12 | 19 | 20 |
| Unknown | Phen-Gen | 43 | 61 | 73 | 92 | 96 | 96 |
| Unknown | Phenotype only | 8 | 21 | 32 | 11 | 27 | 56 |

**Supplementary Table 3**
**Performance when only ESP is used as MAF filter**
Performance in simulated patients with OMIM-listed disease symptoms and HGMD reported variants using only ESP common variants (minor allele frequency >1%) for filtration. The results when all three databases (ESP, dbSNP, 1000 Genomes) are used to define common variants are copied from Table 1 to indicate a drop in performance. The data shows that using all three databases to define common variants, confers a 13-21% advantage.

|  |  | Dominant | | | Recessive | | |
|---|---|---|---|---|---|---|---|
|  | **MAF filtration database** | **Top gene (%)** | **Top 5 genes (%)** | **Top 10 genes (%)** | **Top gene (%)** | **Top 5 genes (%)** | **Top 10 genes (%)** |
| Missense + Nonsense | ESP | 69 | 81 | 84 | 76 | 92 | 94 |
|  | All 3 | 82 | 89 | 91 | 97 | 98 | 98 |
| Splice site | ESP | 69 | 80 | 82 | 73 | 86 | 86 |
|  | All 3 | 80 | 85 | 85 | 87 | 87 | 87 |
| Indels | ESP | 66 | 83 | 83 | 73 | 94 | 95 |
|  | All 3 | 80 | 88 | 94 | 97 | 98 | 98 |
| Combined | ESP | 68 | 82 | 84 | 75 | 92 | 94 |
|  | All 3 | 81 | 88 | 91 | 96 | 97 | 97 |

**Supplementary Table 4**

**Performance in diseases with compound heterozygous inheritance pattern**

Performance for compound heterozygous mutation pairs in coding and genomic regions is depicted. The table entries reflect the percentage of simulations in each category with the indicated result. The results are similar to the 96% observed for recessive simulations in Table 1 for coding variants.

|  | Coding | | | Genomic | | |
| --- | --- | --- | --- | --- | --- | --- |
|  | Top gene (%) | Top 5 genes (%) | Top 10 genes (%) | Top gene (%) | Top 5 genes (%) | Top 10 genes (%) |
| Phen-Gen | 97 | 98 | 98 | 30 | 59 | 62 |
| Genotype only | 25 | 98 | 98 | 8 | 16 | 35 |

**Supplementary Table 5**
**Robustness to symptomatic heterogeneity**

Performance in 44 disorders with variable symptoms is depicted. There were no indels reported for recessive disorders and only one genomic variant. Hence simulations for these variants were omitted. The table entries reflect the percentage of simulations in each category with the indicated result. This is a 5-7% drop off in performance in comparison to known diseases in Table 1 (92% for dominant and 96% for recessive). These results highlight Phen-Gen's robustness to symptomatic heterogeneity.

|  |  | Dominant | | | Recessive | | |
|---|---|---|---|---|---|---|---|
|  |  | Top gene (%) | Top 5 genes (%) | Top 10 genes (%) | Top gene (%) | Top 5 genes (%) | Top 10 genes (%) |
| Phen-Gen | Missense & Nonsense | 88 | 95 | 97 | 90 | 95 | 96 |
|  | Splice site | 86 | 89 | 89 | 86 | 88 | 89 |
|  | Indels | 87 | 91 | 94 |  |  |  |
|  | Combined | 87 | 92 | 95 | 89 | 94 | 95 |

**Supplementary Table 6**
**Performance in simulated patients for regulatory predictor**
The performance in simulated patients with OMIM listed disease symptoms and HGMD regulatory disease causal variants is shown. Phen-Gen's genomic predictor was used for these simulations. The table entries reflect the percentage of simulations in each category with the indicated result.

|  | Dominant | | | Recessive | | | Combined (dominant + recessive) | | |
|---|---|---|---|---|---|---|---|---|---|
|  | Top gene (%) | Top 5 genes (%) | Top 10 genes (%) | Top gene (%) | Top 5 genes (%) | Top 10 genes (%) | Top gene (%) | Top 5 genes (%) | Top 10 genes (%) |
| **Regulatory** | | | | | | | | | |
| Phen-Gen | 58 | 64 | 65 | 40 | 61 | 72 | 49 | 62 | 69 |
| Genotype only | 1 | 2 | 2 | 34 | 51 | 62 | 19 | 29 | 34 |
| Phenotype only | 31 | 43 | 54 | 5 | 11 | 26 | 17 | 26 | 39 |

# Supplementary Table 7
## Performance in real patients

The table reflects performance of Phen-Gen for recessive and X-linked implicated variants in real familial data[5]. The trio IDs correspond to the family identifiers in ref 2. Similarly the gene classification reflects prior knowledge of the gene's involvement in intellectual disability as defined in the original publication. Phen-Gen employs a 1% MAF cutoff whereas the original publication removed all variants reported to dbSNP or observed in their in-house database. For the latter screen all variants common amongst the families was employed. The performance using both these cutoff is depicted.

|  |  |  | Phen-Gen's Rank | |
|---|---|---|---|---|
|  |  |  | Inclusive of dbSNP MAF<1% | Filtration criteria from ref. 2 |
| Trio ID | Gene | Classification |  |  |
| 4 | *FANCB* | Unknown | 18 | 3 |
| 4 | *PDHA1* | Known | 5 | 2 |
| 4 | *GUCY2F* | Unknown | 4 | 1 |
| 16 | *ENOX2* | Unknown | 2 | 1 |
| 18 | *ARHGEF9* | Known | 8 | 1 |
| 25 | *GPM6B* | Unknown | 7 | 1 |
| 41 | *ARHGEF9* | Known | 1 | 1 |
| 42 | *DDX26B* | Unknown | 13 | 3 |
| 72 | *PDZD11* | Unknown | 3 | 3 |
| 93 | *TRPC5* | Candidate | 4 | 1 |
| 12 | *SYCP2L* | Unknown | 2 | 1 |
| 12 | *VPS13B* | Known | 14 | 4 |
| 12 | *C8orf59* | Unknown | 4 | 2 |
| 12 | *PRUNE2* | Unknown | 7 | 3 |
| 24 | *PCNT* | Known | 7 | 2 |
| 70 | *IQGAP2* | Unknown | 6 | 1 |

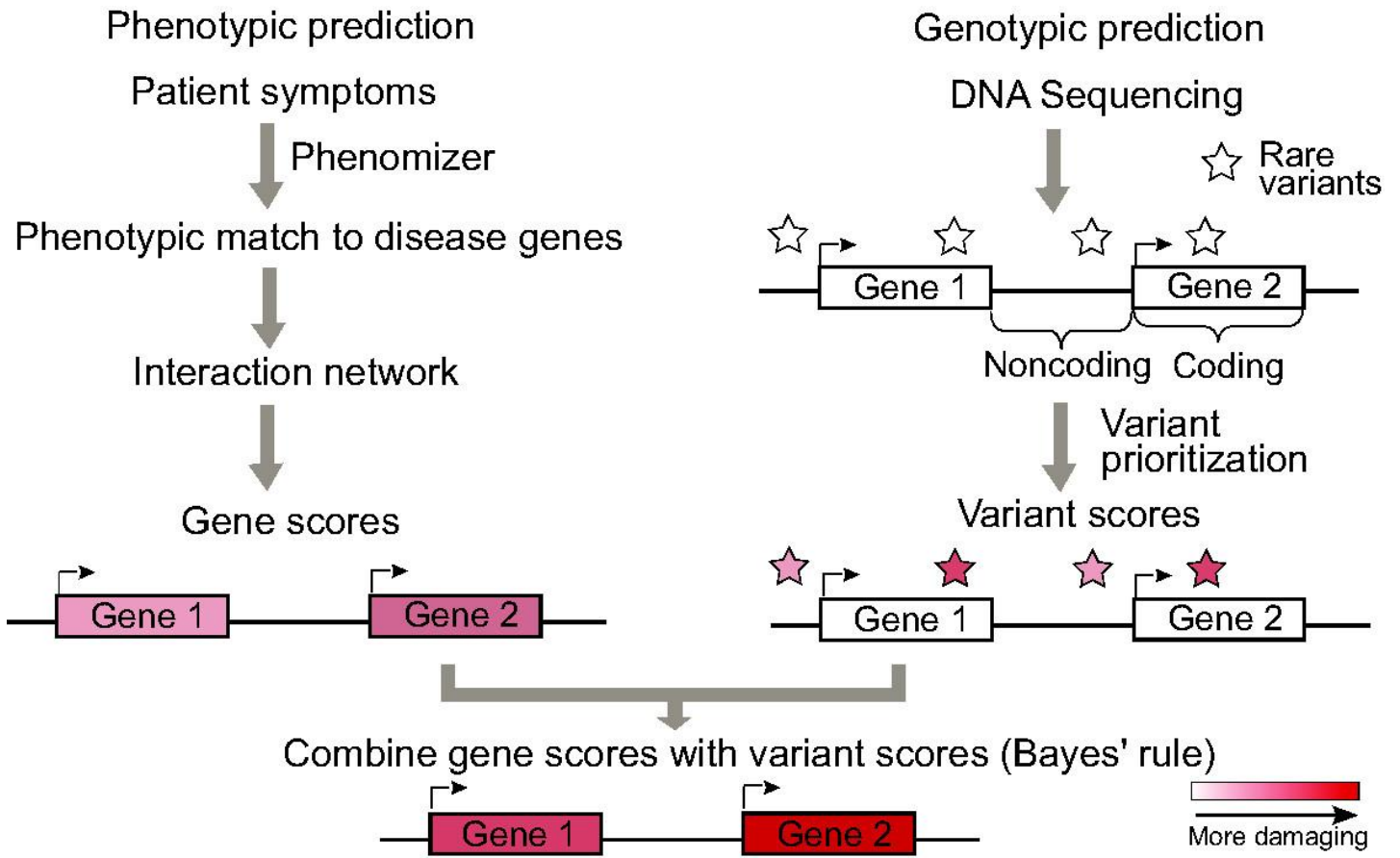

**Supplementary Figure 1**

Overall Workflow.

Patient disease symptoms are matched against known disorders and the probability of a symptomatic match is assigned to genes implicated for the respective disorder. These probabilities are permeated to known gene associates using a random walk with restart on the interaction network. In parallel the patient's sequencing data is analyzed and the damaging impact of each variant estimated and pooled within genes. These two predictions are combined to implicate the gene(s) involved.

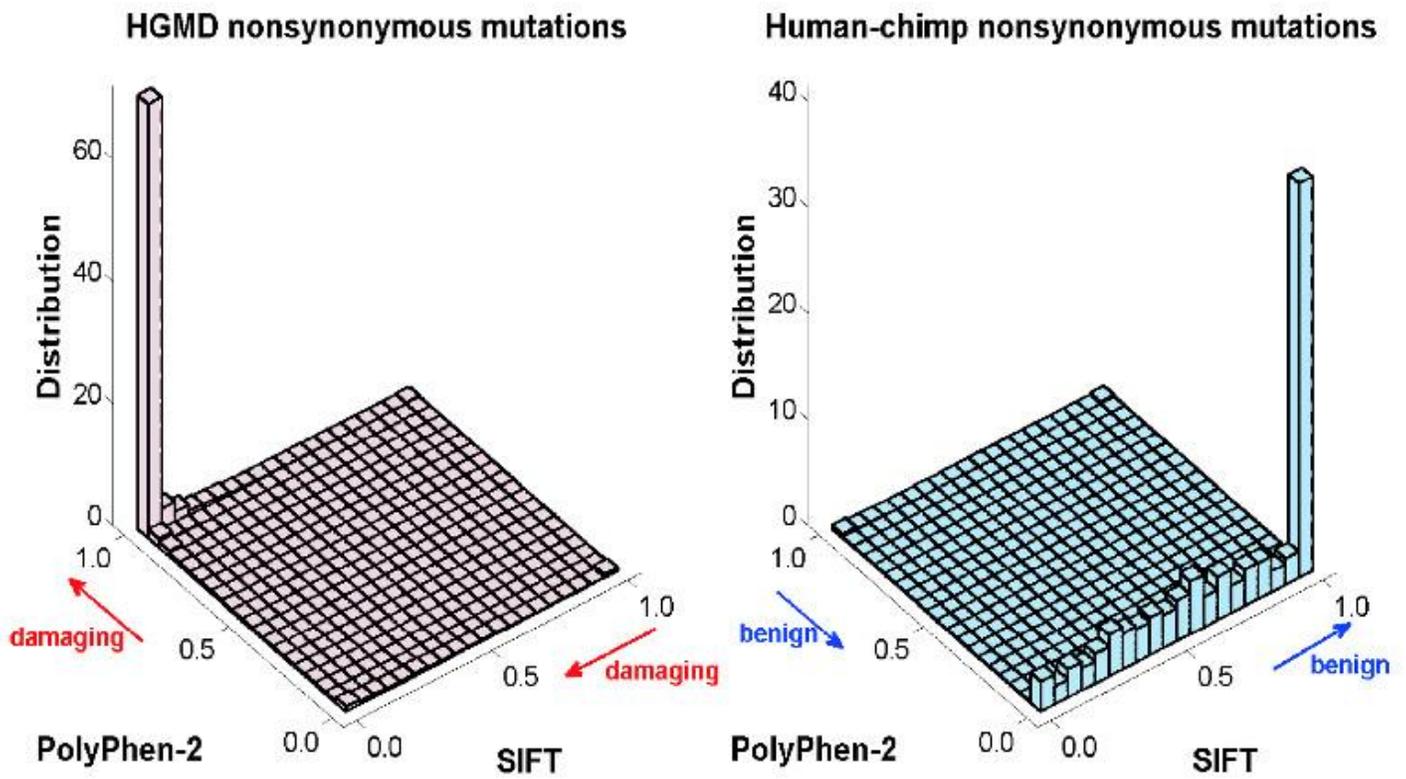

**Supplementary Figure 2**

Distribution of SIFT and PolyPhen-2 scores for damaging and benign nonsynonymous mutations.

The distribution of SIFT and PolyPhen-2 scores for HGMD-reported damaging nonsynonymous mutations and neutral nonsynonymous fixed substitutions inferred from human-chimp alignment are shown. The plots indicate general agreement between the two methods.

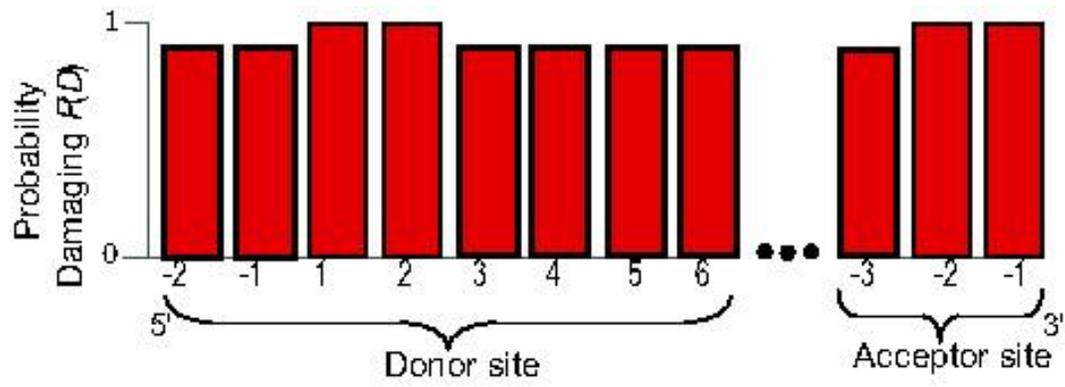

**Supplementary Figure 3**

Deleteriousness predictions around splice site.

The figure depicts the probability of deleteriousness around donor and acceptor sites for splice site mutations.

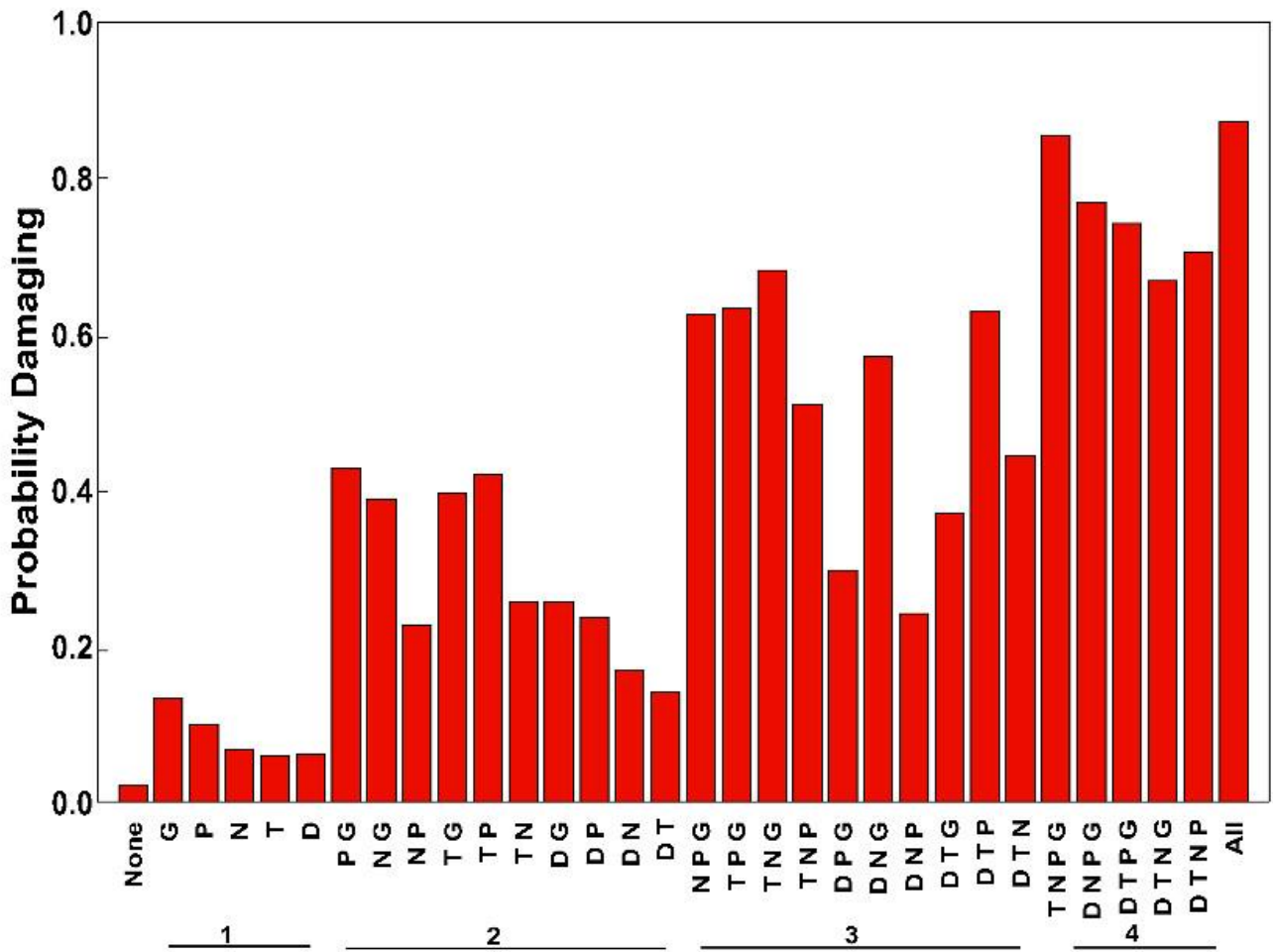

**Supplementary Figure 4**

Probability of deleteriousness using the genomic predictor

The figure illustrates the predicted deleteriousness of different combination of five annotations: GERP++ (G), PhyloP (P), near-genic (N), transcription factor binding sites (T), and DNASE hypersensitive sites (D). The predictions are binned according to the number of annotations (shown on the x-axis). Each bin is further canonically sorted based on the fore mentioned order of annotations.

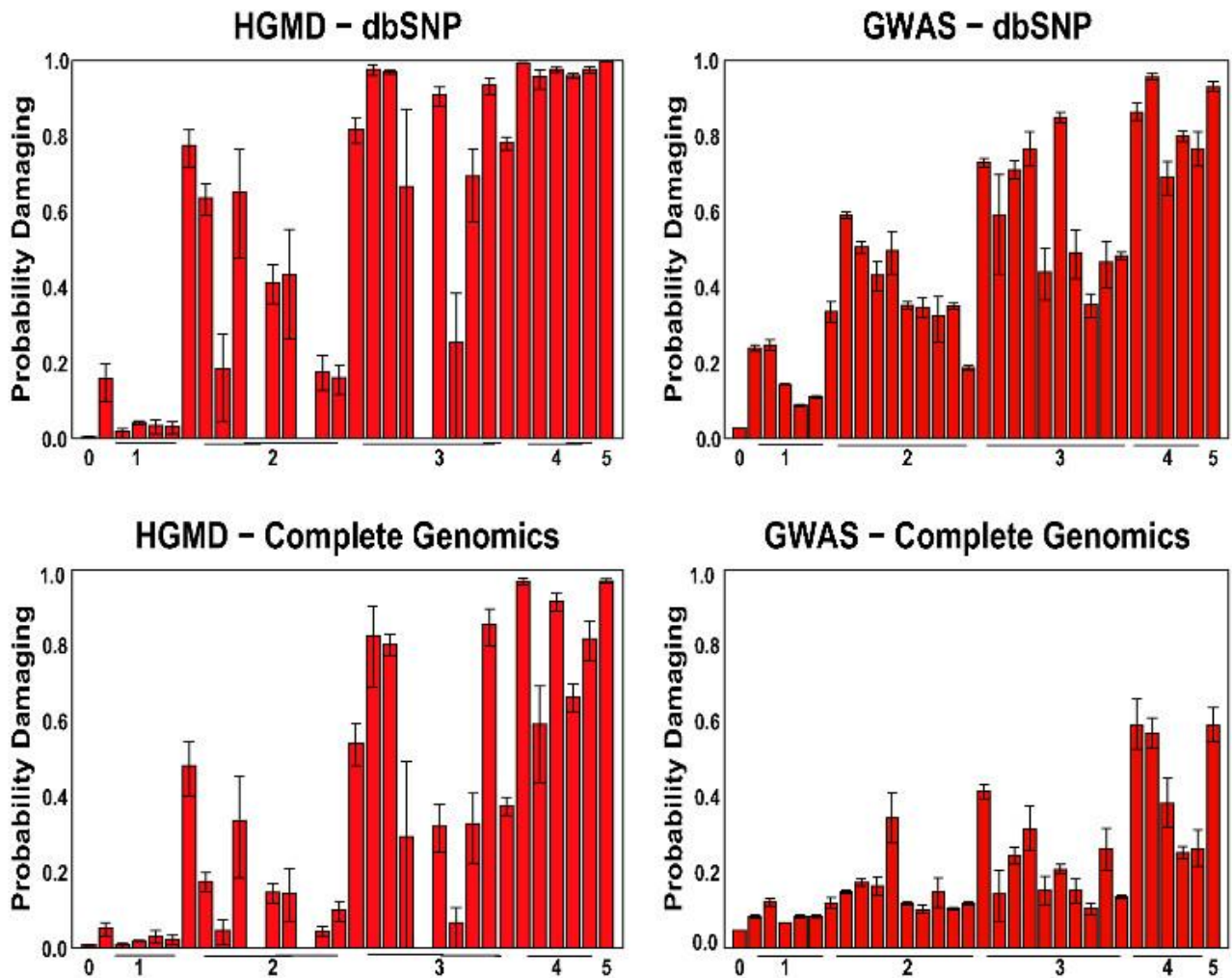

**Supplementary Figure 5**

Confidence intervals for positive and benign mutation set combinations.

The 90% confidence intervals of different combination of genomic annotations are shown. The order from Figure S5 is maintained. With the four sub-figures representing combinations of the two positive sets (HGMD and GWAS) and the two neutral sets (common variation in dbSNP and Complete Genomics MAF>0.30), respectively.

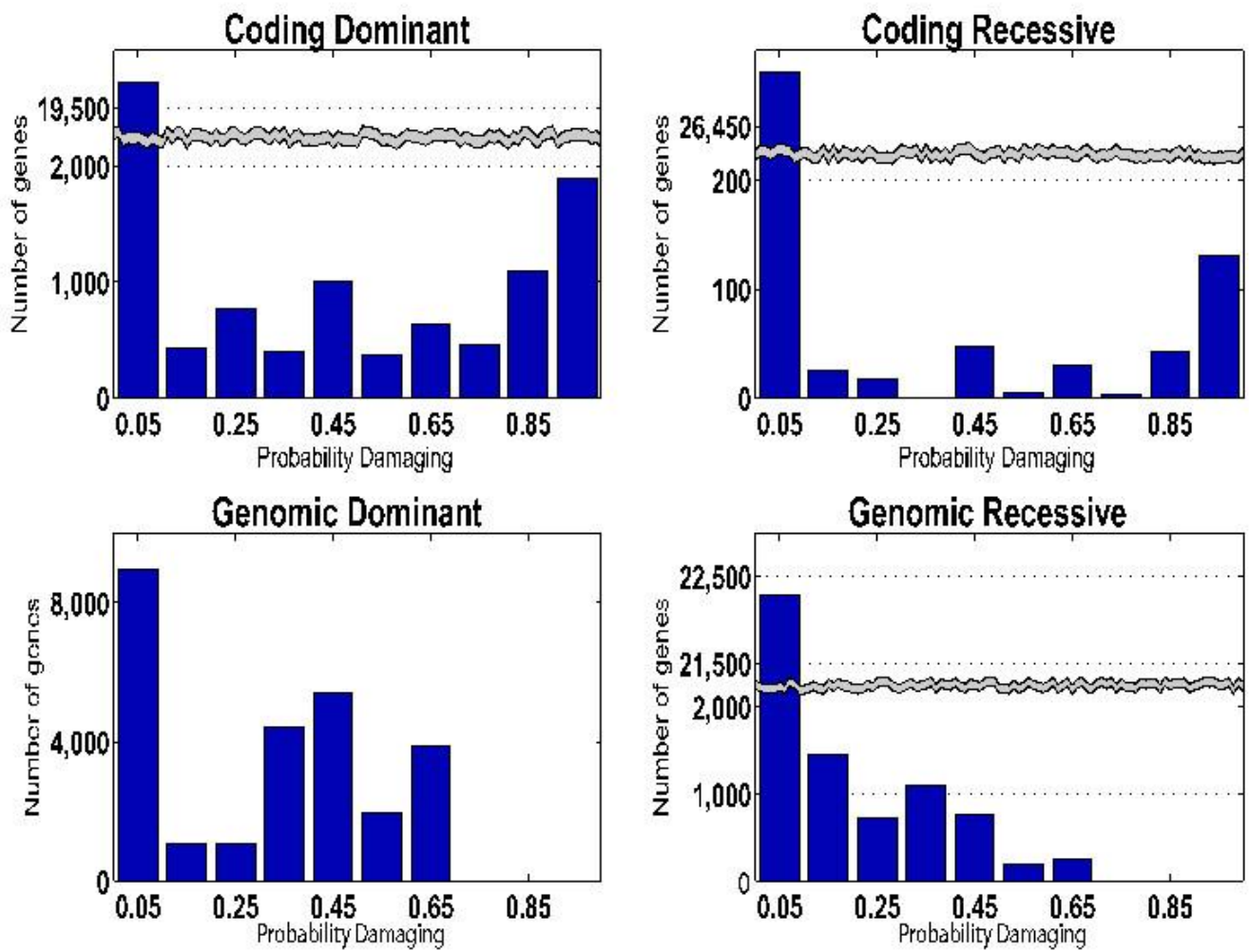

**Supplementary Figure 6**

Histogram of null distribution of deleteriousness of genes.

The top 1 percentile of damaging variants in each gene is shown. The histogram of this null distribution cutoff for all genes under dominant and recessive inheritance pattern for coding and genomic predictors is shown. Most genes do not harbor any putative damaging variants and hence the distributions are dominated by the left most bar; which has been truncated for better visual representation.

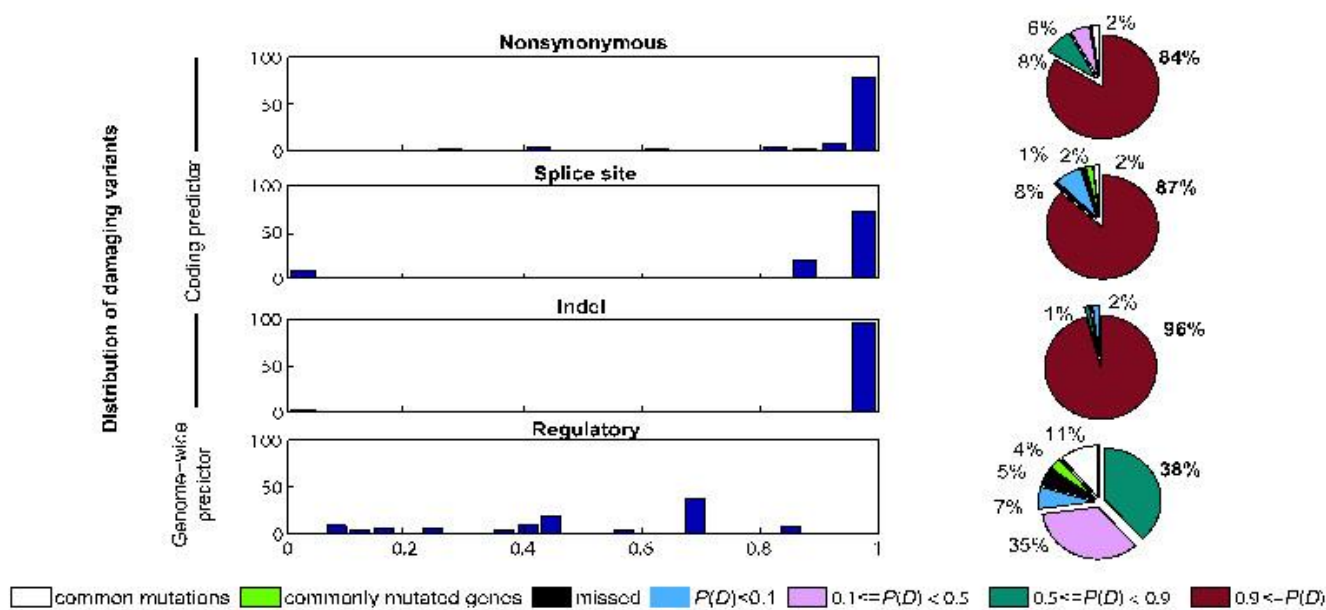

**Supplementary Figure 7**

Performance of variant predictors

The distribution of damaging probabilities assigned to different classes of HGMD variants is shown. The top three panels employ the coding predictor. A genomic predictor was used for the bottom panel and applied to noncoding regulatory variants. The histograms depict the distribution of the scored variants. The pie charts on the right explicate the distribution of omitted and predicted variants in each category. Common variants (light blue) were observed in 1000 Genomes, ESP, or dbSNP with MAF ≥ 0.01. Commonly mutated genes indicate that the variants failed to exceed the null distribution of the respective gene (turquoise). Missed indicates that the variant eluded our regions of interest (dark blue).

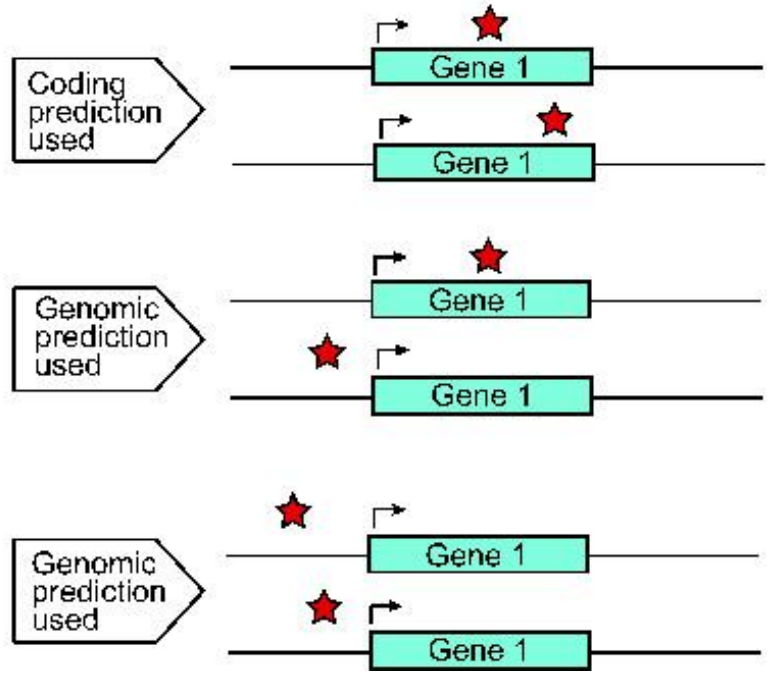

**Supplementary Figure 8**

Prediction of heterozygous variants.

The figure depicts how compound heterozygous variants are evaluated. When both damaging variants reside within the coding region, the coding predictor is used to estimate the damaging impact of these variants. In cases when one or both variants lay outside the exon boundaries, both variants are evaluated using the genomic predictor.

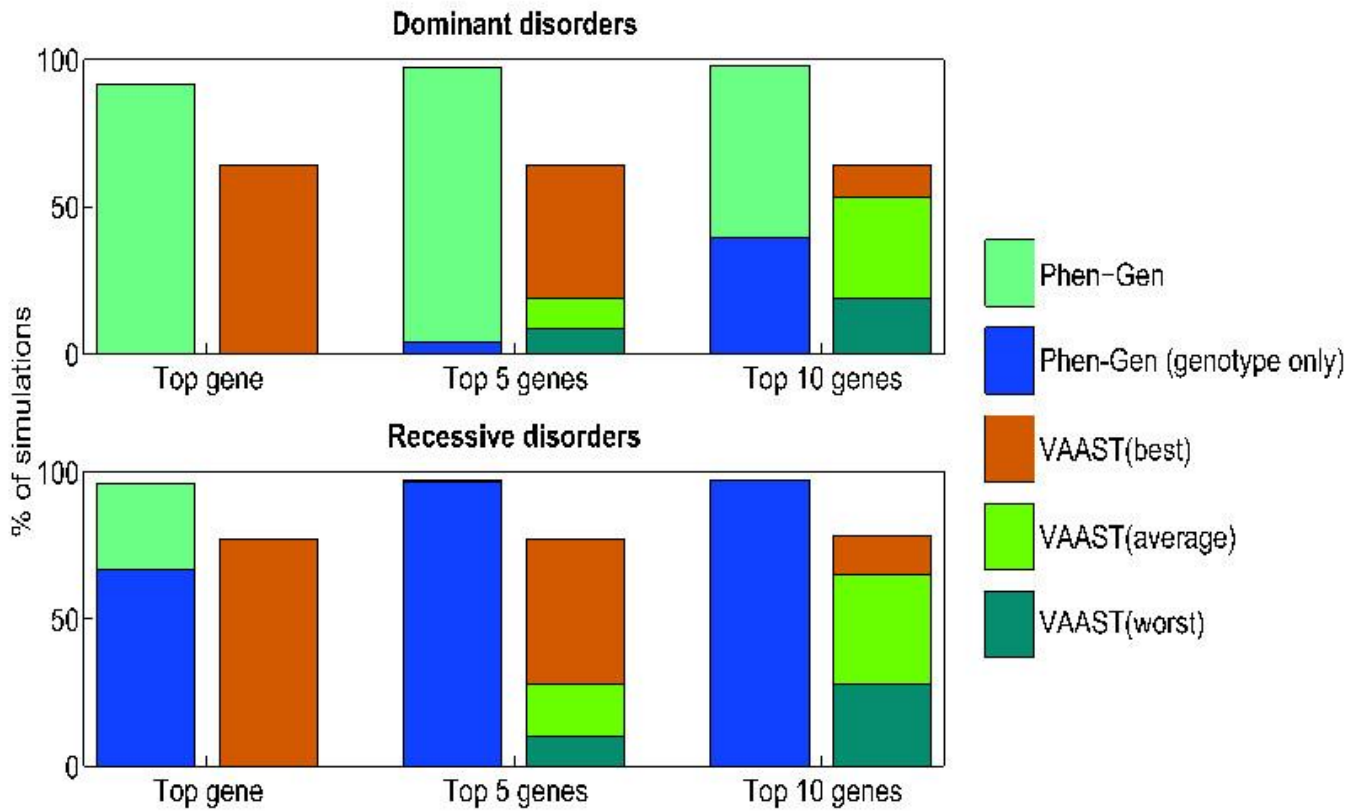

**Supplementary Figure 9**

Phen-Gen and VAAST comparison for phenotypically heterogeneous disorders.

The comparison of Phen-Gen and VAAST in simulations using 44 phenotypically heterogeneous disorders and nonsynonymous mutations in HGMD is shown. In both panels the ability of both methods to narrow down the true gene search within 1, 5 and 10 genes is depicted. For Phen-Gen, the bar is split into the predictive power based on genotypic prediction and the added advantage gained from disease symptoms. VAAST only uses the genomic data and assign multiple genes the same rank at the top of the order. For a fair comparison, the true gene was assigned the worst, average and best rank among similarly ranked peers. The three components of the bar reflect the performance across these scenarios.